\documentclass[12pt]{iopart}
\usepackage{amsfonts}
\usepackage{amssymb}
\usepackage{color,graphicx}

\usepackage{iopams}
\begin{document}

\title[Tunable wideband bandstop acoustic filter]
{Tunable wideband bandstop acoustic filter based on 2D multi-physical phenomena periodic systems}

\author{V. Romero-Garc\'ia}
\ead{virogar1@mat.upv.es}
\address{Instituto de Ciencia de Materiales de Madrid, Consejo Superior de Investigaciones Cient\'ificas}
\address{Centro de Tecnolog\'ias F\'isicas: Ac\'ustica, Materiales y Astrof\'isica, Universidad Polit\'ecnica de Valencia}
\author{J.V. S\'anchez-P\'erez}
\address{Centro de Tecnolog\'ias F\'isicas: Ac\'ustica, Materiales y Astrof\'isica, Universidad Polit\'ecnica de Valencia}
\author{L.M. Garcia-Raffi}
\address{Instituto Universitario de Matem\'atica Pura y Aplicada, Universidad Polit\'ecnica de Valencia}

\begin{abstract}
  The physical properties of a periodic distribution of absorbent resonators is
  used in this work to design a tunable wideband bandstop acoustic filter. Analytical
  and numerical simulations as well as experimental validations show that the control
  of the resonances and the absorption of the scatterers along with their periodic
  arrangement in air introduces high technological possibilities to control noise.
  Sound manipulation is perhaps the most obvious application of the structures presented
  in this work. We apply this methodology to develop a device as an alternative to the conventional
  acoustic barriers with several properties from the acoustical point of view but also with
  additional aesthetic and constructive characteristics.

\end{abstract}
\vspace{2pc}
\maketitle

\section{Introduction}
The interesting propagation properties of electromagnetic waves
inside an inhomogeneous medium with periodic modulation of its
dielectric properties, were simultaneously emphasized by
Yablonovitch \cite{Yablonovitch87}  and John \cite{John87}. These
periodic systems exhibit ranges of frequencies, related to the
periodicity of the structure, where there is no wave propagation. By
analogy with the electronic band gap in semiconductor crystals,
these ranges of frequencies were called band gaps (BG). Motivated by
these results, an increasing interest in the comparable process of
acoustic wave propagation in inhomogeneous periodic acoustic medium
appeared. Several theoretical works started the analysis of periodic
arrays made of isotropic solids embedded in an elastic background
which was also isotropic \cite{Ruffa92, Sigalas92, Kushwaha93}. By
analogy with the electromagnetic case, these periodic arrangements
present BG for elastic waves and they were called phononic crystals
(PC). It is said that if one of the elastic materials in the PC is a
fluid medium, then PC are called sonic crystals (SC). Several
studies discuss the similarities and differences between them
\cite{Sigalas94, Economou93}. Since the acoustical properties of SC
were measured in a minimalist sculpture\cite{Martinez95}, a great
research interest, both experimental and theoretical, have been
emphasized on the existence of complete elastic/acoustic BG, opening
possibilities to interesting applications such as elastic/acoustic
filters, noise control, improvements in the design of transducers,
as well as for the study of pure physics phenomena.

The possibility to manipulate the sound by means of SC motivated the
idea of using these periodic acoustic media as attenuation devices
as, for example, an alternative to noise barriers\cite{Sanchez02}.
However, under the acoustical point of view, the mere existence of
the BG is not sufficient to use SC as noise barriers because both
the size and position of these BG depend on several factors such as
the angle of incidence of the wave or the arrangement of the
scatterers. To avoid these problems, some strategies to increase the
acoustical properties of SC have been developed in last few years.
First, some authors have studied new and more efficient arrangements
of scatterers out of the classical crystalline ones, as
quasi-crystals \cite{Lai02}, quasi-ordered structures
\cite{Romero09, Herrero09} or quasi-fractal arrangements
\cite{Castineira10}. Another strategy is the use of scatterers with
additional properties. This last one can reduce the angular
dependence of the attenuation achieved by the periodic arrangement
and increases both the level and the range of the attenuated
frequencies. In this sense, some authors have designed rigid
scatterers combined with absorbent materials in which both the
multiple scattering and the absorption effects appear in the
periodic arrangement \cite{Umnova06}. Moreover the study of SC built
with scatterers with resonant effects has been done, obtaining good
results especially at frequencies below the BG \cite{Movchan04,
Hu05}. Several works have been very recently developed mixing these
effects in order to increase the properties of the SC as noise
barriers \cite{Sanchez10, Krynkin10}.

In this work we present a wideband bandstop acoustic filter for the
range of the audible frequencies based on the combination of
scattering, resonances and absorption in a 2D periodic array of
scatterers in air. The building blocks of these systems are
scatterers made of a combination of rigid, absorbent and resonant
cavities. \ref{fig:scatterer}A shows the transversal section of the
scatterer considered in this work. A cylindrical rigid wall is
covered by a porous material. A slit is made all along the cylinder
in such a way the shape of the scatterer presents a resonant cavity,
thus these scatterers are based on the concept of split ring
resonator (SRR) introduced by Pendry \textit{et al.} in 1999
\cite{Pendry99} in the
context of the electromagnetism.

The square periodicity of the 2D system is shown in
\ref{fig:scatterer}B. This SC is characterized by two parameters:
the lattice constant $a$, separation between scatterers, and the
filling fraction $ff$, the volume occupied by the scatterers in the
SC. For this periodicity, the main directions of symmetry are
$\Gamma$X (0$^\circ$) and $\Gamma$M (45$^\circ$).In this system, the
periodic distribution of the rigid properties of the scatterers
leads to the phenomenon of the BG, the absorbent material of each
scatterer gives a threshold of attenuation in the audible
frequencies, and finally the resonant cavities introduce attenuation
in relation with the resonances. But all these effects can be fitted
by introducing high technology procedures in the design of devices
to control the noise. We have studied both theoretically and
experimentally and progressively the acoustic effect of each
component of the scatterer on the frequency response of the whole
system. From the theoretical point of view we analyse the effect of
each phenomenon by means of several theoretical and numerical
techniques.

\begin{figure}[hbt]
\begin{center}
(A)\hspace{6cm}(B) \caption{\label{fig:scatterer} Schematic view of
both the scatterer and the SC configurations of the device
presented. (A) The transversal view of the 1 m long scatterer
considered. The inner radius is $r_{int}=0.095$ m, the exterior
radius is $r_{ext}=0.14$ m, the aperture $L=0.02$ m, the rigid wall
thickness $\Delta_r=0.005$ m and the absorbent covering thickness
$\Delta_a=0.04$ m. (B) The size of the designed device is $4a\times
4a$. The scatterers are arranged in a square array with lattice
constant $a=0.33$ m. For this square periodicity the main directions of symmetry of the SC are $\Gamma$X (0$^\circ$) and $\Gamma$M (45$^\circ$).} 
\end{center}
\end{figure}

Thus, the infinite periodic distribution of an array of rigid
scatterers is analyzed by means of the dispersion relation using the
Plane Wave Expansion method (PWE). This method allows to understand
the transmission properties of the wave inside the crystalline
systems.
However, in the real situations the finite sample effects have to be
taken into account. In this sense, the Multiple Scattering Theory
(MST) is introduced as a methodology to study the
scattering in a finite distribution of cylindrical scatterers.
The methodology is developed and used for two cases, rigid and
absorbent scatterers, in order to compare and separate the
contributions of both the absorbent material and the effect of the
periodicity of the array in the total attenuation obtained. The
results are compared with those obtained using PWE. The case of the
resonant effect, due to the complexity of the shape of the
scatterers because of the existence of the split all along the
cylinders, has been analyzed numerically using the Finite Elements
Method (FEM).
The combined effect of scattering, resonances and absorption in the
definitive SC device is studied both numerically and experimentally.
Finally, the main applications of such a structures and the main
conclusions of the work are
summarized in last Sections.

\section{Theoretical and numerical methods}
A wide range of mathematical techniques are now available for the
resolution of problems involving the interaction of waves with
arrays of scatterers. This Section draws together the methods used
in this work: PWE, MST and FEM.

In order to show the main results obtained by each theoretical
method explained in this Section, we have applied them to study the
propagation properties of a SC made of rigid or absorbent
cylindrical scatterers with $r=0.1$ m, similar to that shown in
\ref{fig:scatterer}A but with $L=0$. These scatterers are arranged
in a square array with $a=0.33$ m. The filling fraction of the array
is $ff=\frac{\pi r^2}{a^2}=0.2885$ (28.85 \%). This previous study
allow us to understand the acoustic properties of the definitive
designed device.

\subsection{Analytical models}

\subsubsection{Eigenvalue problem: plane wave expansion (PWE)}
\label{sec:PWE} The dispersion relation of a 2D SC made of rigid
cylindrical scatterers can be obtained by means of the solution of
the eigenvalue problem using the PWE. These bands structures are a
good representation to understand the dispersion of waves inside
such a crystalline systems. This method is valid only for infinite
periodic crystals.

By analogy to electron waves in a crystal, waves transmission inside
periodic systems should be described using the bands theory. This
idea was first introduced in 1987 \cite{Yablonovitch87, John87} and
then the concepts of Bloch waves, dispersion relations, Brillouin
zones and so on, can be applied to the case of all kind of periodic
systems: photonic, phononic and sonic crystals \cite{Yablonovitch89,
Meade92, Economou93, Kushwaha94}. PWE is a methodology that uses the
periodicity of the system and the Bloch theorem to solve the wave
equation, obtaining a simple eigenvalue problem
relating to the wave vector and the frequency of the incident wave.

Propagation of sound is described by the equation
\begin{eqnarray}
\frac{1}{\rho c^2} \frac{\partial^2 p}{\partial
t^2}=\nabla\left(\frac{1}{\rho}\nabla p \right)
\label{eq:acoustic}
\end{eqnarray}
where $c$ is the sound velocity, $\rho$ is the density of the
medium and $p$ is the pressure.

In this Section, an array of straight, infinite cylinders made of an
isotropic solid $A$, embedded in an acoustic isotropic background
$B$ has to be considered. There is translational invariance in the
direction $z$ parallel to the cylinders and the system has 2D
periodicity in the transverse plane. Thus, this assumption implies
an infinite medium in order to obtain its propagation properties. By
using this periodicity, it is possible to expand the properties of
the medium in Fourier series,
\begin{eqnarray}
\sigma=\frac{1}{\rho(\vec{r})}=\sum_{\vec{G}}\sigma_{\vec{k}}(\vec{G})e^{\imath \vec{G}\vec{r}} \label{eq:sigma},\\
\eta=\frac{1}{B
(\vec{r})}=\sum_{\vec{G}}\eta_{\vec{k}}(\vec{G})e^{\imath
\vec{G}\vec{r}}\label{eq:eta},
\end{eqnarray}
where $\vec{G}$ is the 2D reciprocal-lattice vector, $\vec{G}=(G_1, G_2, G_3)=(2\pi m/a_1,2\pi n/a_2, 0)$ for square periodicity, and
$B(\vec{r})=\rho(\vec{r})c(\vec{r})^2$ is the bulk modulus. The pressure $p$ can be obtained by applying the Bloch theorem and harmonic temporal
dependence,
\begin{eqnarray}
p(\vec{r},t)=e^{\imath (\vec{k}\vec{r}-\omega
t)}\sum_{\vec{G}}p_k(\vec{G})e^{\imath \vec{G}\vec{r}}.
\label{eq:pressure}
\end{eqnarray}


Using Equations (\ref{eq:sigma}), (\ref{eq:eta}),
(\ref{eq:pressure}) and (\ref{eq:acoustic}) we obtain
\cite{Kushwaha94}
\begin{eqnarray}
\sum_{\vec{G'}}\left((\vec{k}+\vec{G})\sigma_k(\vec{G}-\vec{G'})(\vec{k}+\vec{G'})-\omega^2\eta_{\vec{k}}(\vec{G}-\vec{G'})\right)p_{\vec{k}}(\vec{G'})=0.
\label{eq:eigenproblem}
\end{eqnarray}
For $\vec{G}$ taking all the possible values, equation
(\ref{eq:eigenproblem}) constitutes a set of linear, homogeneous
equations for the eigenvectors $p_{\vec{k}(\vec{G})}$ and the
eigenfrequencies $\omega({\vec{k}})$. Further mathematical formulation as well as an extended method
to solve the inverse problem $k(\omega)$ is shown in reference \cite{Romero10c}.

\begin{center}\begin{table}
\caption{Directions of incidence, ranges of $\vec{k}$, and ranges of
phase changes, $\vec{k}\cdot \vec{R_1}$ and $\vec{k}\cdot
\vec{R_2}$, for each of the segments required to traverse the
boundary of the irreducible first Brillouin zone for square
lattice.}
\begin{center}
\begin{tabular}{c|ccc}
 \hline
  & \multicolumn{3}{|c}{Square lattice}\\
Direction & $\vec{k}$ &  $\vec{k}\cdot \vec{R_1}$ & $\vec{k}\cdot \vec{R_2}$ \\
\hline \hline
$\Gamma$X &  $[(0,0) (0,\pi/a)]$ &  $[0,\pi/a]$ & $[0,0]$\\
XM &  $[(\pi/a,0) (\pi/a,\pi/a)]$ &  $[\pi/a,\pi/a]$ & $[0,\pi/a]$\\
M$\Gamma$ & $[(\pi/a,\pi/a) (0,0)]$ &  $[\pi/a,0]$ &  $[\pi/a,0]$ \\
 \hline
\end{tabular}
\end{center}
\label{tab:k}
\end{table}
\end{center}

By solving the system given in \ref{eq:eigenproblem} for each Bloch
vector in the irreducible area of the first Brillouin, zone the
eigenvalues $\omega^2$ are obtained, and they can be used to
represent the band structures or dispersion relation
$\omega(\vec{k})$ in the periodic system. \ref{tab:k} shows the
directions of incidence, ranges of $\vec{k}$, and ranges of phase
changes, $\vec{k}\cdot \vec{R_1}$ and $\vec{k}\cdot \vec{R_2}$, for
each of the segments required to traverse the boundary of the
irreducible first Brillouin zone for the square lattice.

In \ref{fig:Figure_2_def}A one can observe the bands structure of
the selected SC. The frequencies are represented versus the Bloch
vector in the first Brillouin zone which is in relation with the
incident direction of the wave. Each black line represents a
propagating band, i.e., allowed modes inside the periodic structure.
Note that for the $\Gamma X$ direction there is a range of
frequencies (marked with horizontal dashed lines) in which there is
no propagation. This forbidden frequency range for a specific
incident direction is known as a pseudogap. In this angle of
incidence and for this frequencies the SC works as an acoustic
filter.

\begin{figure}[hbt]
\begin{center}
\caption{ Theoretical propagation properties of a SC made of rigid
or absorbent scatters obtained with the three methods used in this
work. (A) Bands structure (or dispersion relation) of a SC made of
rigid scatterers with $a=0.33$ m and $r=0.1$ m. Black lines
represent the analytical predictions calculated using PWE. Red dots
show the numerical predictions calculated using FEM. Horizontal
dashed lines represent a pseudogap. (B) and (C): corresponding scattering problem calculated for a SC
of size $4a\times4a$ made of rigid or absorbent scatterers respectively. The SC is placed at 0.4 m from the source and centered with respect to the source-receiver line. The site of the receiver is $(2.49,0)$ m. Blue
line (Blue dots) represents the attenuation spectrum for the $\Gamma
X$ direction predicted by MST (FEM). Red dashed line (Red open
circles) represents the attenuation spectrum for the $\Gamma M$
direction predicted by MST (FEM).} \label{fig:Figure_2_def}
\end{center}
\end{figure}

\subsubsection{Scattering problem: Multiple Scattering Theory (MST)} \label{MST}
In real situations the finite sample effects have to be taken into
account. Multiple scattering Theory gives the possibility to study
the scattering problem in these finite structures. Motivated by the
work of Tournat \textit{et al.} \cite{Tournat04}, in which
scatterers with a mesoscopic scale much larger than the microscopic
scale are placed in a porous medium as host material, we briefly
present here the multiple scattering of a 2D array of scatterers
made of a rigid core covered with a layer of absorbing materials.
The inner rigid core is a cylinder with radius $r_{in}$, and the
covering of the absorbing material has a thickness $t$, so that the
external radius is $r_{ext}=r_{in}+t$. Other works \cite{Umnova06}
show similar methodology as in this Section.

Absorbing materials usually present a complex impedance,
$Z_c(\omega)$ and complex propagation constant, $k_c(\omega)$, both
being frequency dependent. Therefore we will consider two different
boundary conditions: one corresponding to the rigid core and the
other across the interface between the absorbing covering of the
scatterer and the surrounding medium. The formalism developed in
this Section shows a general procedure,
independently from the expressions of $Z_c$ and $k_c$ used to model the absorbing material. The porous
material will be described by means of the Delany-Bazely model, then
\begin{eqnarray}
Z_c(\nu)=1+0.0571\left(\frac{\rho_0 \nu}{R}\right)^{0.754}-\imath
0.087\left(\frac{\rho_0 \nu}{R}\right)^{-0.732},\\
k_c(\nu)=k_0\left(1+0.0928\left(\frac{\rho_0
\nu}{R}\right)^{-0.7}+\imath 0.189\left(\frac{\rho_0
\nu}{R}\right)^{-0.597}\right),
\end{eqnarray}
where $\rho_0$, $c_0$ represent the density and the sound velocity of the
air respectively; $k_0$ is the wave number of the wave propagating
in air; $R$ is the flow resistivity; and $\nu$ is the frequency of
sound ($\omega=2\pi\nu$). Here $R=23000$ Pa s m$^{-2}$ corresponds to the woollen felt covering. In order to avoid non-physical results the signs of the imaginary part of $Z_c$ and $k_c$ have been selected properly. This improves the results shown in reference \cite{Umnova06}. This model presents some ranges of applicability dependent on both the resistivity of flow and the frequency. For the range of values of these parameters, the Delany-Bazley model works properly in the range of frequencies 186$<\nu<$18700 Hz, which is basically the range of frequencies in the audible range (20-22000 Hz).


An acoustic source transmitting white noise is placed at point
$\vec{r_s}$, located at some distance from the
system of scatterers. For the sake of simplicity, without
compromising generality, we approximate the acoustic source as a
line source located at origin, i. e. $\vec{r_s}=\vec{0}$. The
acoustic wave emitted by such a source follows the Equation in
cylindrical coordinates:
\begin{eqnarray}
P_0(\vec{r})=\imath \pi H_0(kr),
\end{eqnarray}
where $H_0$ is the zero$-th$ order Hankel function of the first kind.
The solution represents a line source located at origin.

We consider $N$ straight scatterers located at
$\vec{r_i}=(r_i,\theta_i)$ with $i=1, 2,\ldots,N$
to form either a regular lattice or a random array perpendicular
to the $x-y$ plane. The scatterers are parallel to the $z-$axis, then
since the boundary conditions and the geometry do not change with
$z$, the problem can be reduced to two uncoupled problems for the
scalar Helmholtz equation. The final wave reaches a receiver
located at $\vec{r_r}$ and it is formed by the sum of the direct wave from the source and the
scattered waves from all the scatterers.

The incident wave over $i$-th scatterer at a point $\vec{r}$ outside the scatterer is:
\begin{eqnarray}
P^i_{in}(\vec{r})=\sum_{n=-\infty}^{\infty}B^i_n J_n(k|\vec{r}-\vec{r_i}|)e^{\imath n\phi_{\vec{r}-\vec{r_i}}}.
\end{eqnarray}
On the other hand, the scattered wave produced by $i$-th scatterer at a point $\vec{r}$ outside the scatterer is:
\begin{eqnarray}
P^i_{sc}(\vec{r},\vec{r_i})=\sum_{n=-\infty}^{\infty}\imath \pi A^i_n H^{(1)}_n(k|\vec{r}-\vec{r_i}|)e^{\imath n\phi_{\vec{r}-\vec{r_i}}},
\end{eqnarray}
where $H_n$ is the n$-th$ order Hankel function of the first kind, and $J_n$ is the n$-th$ order Bessel function of the first kind.

The wave transmitted within the absorbing material of
the $i$-th scatterer at a point $\vec{r}$ inside
the absorbing layer is:
\begin{eqnarray}
P^i_{int}(\vec{r},\vec{r_i})=\sum_{n=-\infty}^{\infty} A^i_n (X_n^i H^{(1)}_n(k_c(\omega )|\vec{r}-\vec{r_i}|)+Y^i_n J_n(k_c(\omega )|\vec{r}-\vec{r_i}|))e^{\imath n\phi_{\vec{r}-\vec{r_i}}},\nonumber \\
\end{eqnarray}
where $X_n^i$ and $Y_n^i$ are two coefficients to be calculated
later with appropriate boundary conditions. Then, the external wave
outside the $i$-th scatterer at a point $\vec{r}$ outside the
scatterer is:
\begin{eqnarray}
P_{ext}(\vec{r},\vec{r_i})=\sum_{n=-\infty}^{\infty} \left(B^i_n J_n(k|\vec{r}-\vec{r_i}|) + \imath \pi A^i_n H^{(1)}_n(k|\vec{r}-\vec{r_i}|)\right)e^{\imath n\phi_{\vec{r}-\vec{r_i}}}.
\end{eqnarray}

Due to the scatterers considered in this Section, the problem
presents two different kinds of boundary conditions. In the walls of
the rigid core, one can consider Neumann boundary conditions.
However, in the absorbing material-host medium interface, one should
consider the continuity of the pressure and the velocity. Thus, the
boundary condition in the rigid wall, $\Gamma_i$, inside $i$-th
scatterer is:
\begin{eqnarray}
\label{eq:neu}
\frac{\partial
P^i_{int}}{\partial n}|_{\Gamma_i}=0
\end{eqnarray}
and the boundary conditions in the external
interface, $\Omega_i$, of the scatterer are,
\begin{eqnarray}
\label{eq:continuous_1}
p^i_{ext}|_{\partial \Omega_i}=p^i_{int}|_{\partial \Omega_i}\\
\label{eq:continuous_2}
\frac{Z_c(\omega) k_c(\omega)}{k_0}\frac{\partial
p_{ext}}{\partial n}|_{\partial \Omega_i
}=\frac{\partial p_{int}}{\partial
n}|_{\partial \Omega_i}
\end{eqnarray}
where $\partial \Omega_i$ is the boundary of the $i$-th scatterer,
$k_0$ is the wave number in the host medium, $k_c(\omega)$ and $Z_c(\omega)$ are the propagation constant and the impedance of the absorbing material of the scatterer $i$.

By applying the boundary condition (\ref{eq:neu}) in $\Gamma_i$, we can obtain a simple relation between coefficients $X_n^i$ and $Y_n^i$:
\begin{eqnarray}
Y^i_n=X_n^i T_n^i\\
T^i_n=-\frac{H'_n(k_c(\omega)r^i_{in})}{J'_n(k_c(\omega)r^i_{in})}.
\end{eqnarray}
where the prime as superscript represents the derivative with respect to the normal of the surface.

Finally, applying the boundary condition (\ref{eq:continuous_2}) at the $\partial \Omega_i$ interfaces, we get:
\begin{eqnarray}
B^i_n=\imath \pi Z^i_n A^i_n,
\end{eqnarray}
where,
\begin{eqnarray}
Z^i_n=-\frac{f(\omega)H'_n(k r^i_{out})-H_n(k r^i_{out})}{f(\omega)J'_n(k r^i_{out})-J_n(k r^i_{out})},\\
f(\omega)=\frac{Z_c(\omega)k_c(\omega)}{k}\frac{H_n (k_c(\omega) r^i_{out})+T^i_n J_n(k_c(\omega)r^i_{out})}{H'_n (k_c(\omega) r^i_{out})+T^i_n J'_n(k_c(\omega)r^i_{out})}.
\end{eqnarray}

Note that the previous equations also reproduce the case of rigid
scatterers. If the absorbing cover is not considered, then
$r_{out}=r_{in}$, $Z_c(\omega)=1$ and $k_c(\omega)=k$ and the
equations obtained are the same as the ones previously obtained by
several authors \cite{Chen01}.

The attenuation spectrum of an arrangement of scatterers is obtained by the representation of the
insertion loss (IL) which is the difference between the sound level
recorded with and without the sample. Note that throughout this work
the IL is calculated as
\begin{equation}
\label{form_IL}
    IL=20\log_{10}\left|\frac{P_0}{P}\right|.
\end{equation}
where $P$ is calculated as
\begin{eqnarray}
P(\vec{r})=\imath\pi H_0(kr)+\sum_{i=1}^N\sum_{n=-\infty}^{\infty} \imath \pi A^i_n H^{(1)}_n(k|\vec{r}-\vec{r_i}|)e^{\imath n\phi_{\vec{r}-\vec{r_i}}},
\end{eqnarray}
being N the total number of scatterers.

\ref{fig:Figure_2_def}B and \ref{fig:Figure_2_def}C show the MST
predictions for the SC defined above and for the case of rigid and
absorbing scatterers. Blue continuous line represents the
attenuation spectrum for the $\Gamma X$ direction and red dashed
line represents the attenuation spectrum for the $\Gamma M$
direction. For the case of the rigid scatterers one can see the
effect of the periodicity of the structure: Blue line shows the
pseudogap at $\Gamma X$ direction between 350 Hz and 630 Hz. This
pseudogap disappears for the $\Gamma M$ direction. These results are
in completely agreement with the predictions of PWE shown in
\ref{fig:Figure_2_def}A. In this structure only the scattering due
to the rigidity of the scatterers is playing a role in the
attenuation process producing a low and angle dependent attenuation
level. In the case of the absorbent scatterers one can observe the
acoustic effect of the absorbent material: the average of the IL
increases in the whole range of the considered frequencies.

\subsection{Finite Elements Method (FEM)}
\label{sec:FEM} Sometimes the geometrical shape of the scatterers or
the concurrence of several effects (multiphysics) are difficult to
solve by means of an analytical method. In these cases numerical
methods seem a good alternative to find solutions. For the problem
we are dealing with, SC made of resonant, absorbent and scattering
building blocks, we have used the Finite Elements Method to obtain
the propagating properties of the system. In this work the
commercial software COMSOL Multiphysics 3.5 is used.

\subsubsection{Eigenvalue problem for rigid scatterers: bounded problem}

For solving the problem using FEM, it is necessary to define the
symmetry, discretize the domain and consider the boundary
conditions. In the boundary of each scatterer both the continuity of
the pressure and the velocity are considered. As we have explained
before, for rigid scatterers we can use both the Neumann boundary
condition and Bloch theorem due to the translational symmetry. The
properties of the Bloch states constrains the solution to a unit
cell with Bloch vectors in the first Brillouin zone. These features
transform the unit cell in a bounded domain to solve the problem
with the next boundary condition at the borders of the unit cell,
\begin{eqnarray}
\label{eq:7} P(\vec{r}+\vec{R})=P(\vec{r})e^{\imath \vec{k}\vec{R}}
\end{eqnarray}
where $k$ is the Bloch vector and it scans the first irreducible
Brillouin zone (see \ref{tab:k} for square periodicity).


Red circles in \ref{fig:Figure_2_def}A show the bands structure
predicted using FEM. One can compare the very good agreement between
the FEM results with the ones obtained using PWE.

\subsubsection{Scattering problem: unbounded problem}
Considering the wave propagation in free space (unbounded acoustic
domain) the assumption that no waves are reflected from infinity is taken.
This is known as the Sommerfeld condition. 
The solutions of exterior Helmholtz problems that satisfy the
Sommerfeld conditions are called radiating solutions. Using FEM it
is only possible to obtain some approximation of the radiating
solutions in unbounded domains by applying some artificial
boundaries in the numerical domain. Several techniques can be used
for this purpose \cite{ihlenburg98}. Among them the perfectly
matched layers (PML) will be briefly presented in this section.

The PML method was introduced by Berenger \cite{Berenguer94} and it is an efficient alternative for emulating
the Sommerfeld radiation condition in the numerical solution of
wave radiation and scattering problems. 
The method was immediately applied to different cases based on the
scalar Helmholtz equation \cite{Harari00} acoustics
\cite{Abarbanel99, Qi98}, elasticity \cite{Basu03}, poroelastic
media \cite{Zeng01}, shallow water waves \cite{Navon04}, other
hyperbolic problems \cite{Lions02}, etc. Here, the interest is
focused on the wave propagation time-harmonic scattering problems in
linear acoustics, i.e., on the scalar Helmholtz equation.

PML consists of a coordinate transformation \cite{Liu99, Collino98}. The transformation is a scaling to complex coordinates so that the new me\-dium becomes selectively dissipative in the direction perpendicular to the interface between the PML and the physical domain. In this work, the PML domain absorbs waves in the coordinate direction $d$ following the next coordinate transformation inside the PML:
\begin{eqnarray}
d'=sign(d-d_0)|d-d_0|^n\frac{L}{\delta D^n}(1-\imath)
\end{eqnarray}
where $L$ is the scaled PML width, $d_0$ is the coordinate of the
inner PML boundary, the width of the PML is $D$ and $n$ is the scaling
exponent. This coordinate transformation is provided in COMSOL 3.5.


In practice, since the PML have to be truncated at a finite
distance of the domain of interest, its external boundary produces
artificial reflections. Theoretically, these reflections have
minor importance due to the exponential decay of the acoustic
waves inside the PML. In fact, for Helmholtz-type scattering
problems, it was proven that the approximate solution
obtained using the PML method exponentially converges to the exact
solution in the computational domain as the thickness of the layer
goes to infinity \cite{Lassas98}. This result was generalized using techniques based on the pole condition \cite{Hohage03}.

\ref{fig:Figure_2_def}B and \ref{fig:Figure_2_def}C show the FEM
predictions for the considered SC made of rigid or absorbing
scatterers respectively. Blue dots (red open circles) represent the
attenuation spectrum for the $\Gamma X$ ($\Gamma M$) direction. The
difference to the results obtained with MST at high frequencies
could be due to the low number of elements in the mesh of FEM.

\section{Laboratory experiment}
All the experimental results shown in this work have been measured
under controlled conditions in an anechoic chamber. A picture of the
designed definite scatterers and a scheme of the anechoic chamber
are shown in \ref{fig:Figure_3_def}. In \ref{fig:Figure_3_def}B
pictures of both the periodic array of these scatterers and of the
robotized acquisition hardware are also shown.
\begin{figure}[hbt]
\begin{center}
\caption{Experimental set up. (A) Picture of the single designed
scatterer. (B) Scheme of the anechoic chamber, and pictures of both
the experimental device designed and the robotized system 3DReAMS.}
\label{fig:Figure_3_def}
\end{center}
\end{figure}

 All the acoustic measurements were received in a
prepolarized free-field microphone 1/2" Type $4189$ B\&K. This
microphone is controled by a 3D Robotized e-Acoustic Measurement
System (3DReAMS), which is a Cartesian robot with three axes (X, Y,
Z) installed in the ceiling of the anechoic chamber. The robot was
designed to sweep the microphone through a 3D grid of measuring
points located at any trajectory inside the echo-free chamber. The
robot includes a rotatory column installed on the ceiling of the
anechoic chamber, where the periodic arrays are hung in a frame.

The National Instruments cards PCI-4474 and NI PCI-7334 were used together with the Sound and Vibration Toolkit and the Order Analysis Toolkit for LabVIEW for both the data acquisition and the motion of the robot. Once the robotized system is turned off and the acoustic source and the microphone are turned on, the microphone acquires the temporal signal. From this temporal signal, one can obtain the power spectra, the frequency response or the sound-level measurement.


\section{Results and discussion}
Recently some authors have specifically constructed acoustic
applications using split ring resonators (SRR) \cite{Hu05} motivated by
the fact that these devices introduce ranges of frequencies related
to the resonant frequency where waves cannot propagate through the
system \cite{Movchan04}. Although authors have usually considered
the SRR as 2D Helmholtz resonators, this approximation needs some
special geometrical restrictions \cite{mechel08}: The thickness of
the walls or both the length and the aperture of the neck of the
resonator have to follow some approximations to be considered as a
Helmholtz resonator. Otherwise, one should solve the scattering
problem of the isolated resonator in order to know the resonant
frequency of the SRR, as we have done in this work. So, once the
resonant frequency of the SRR is known, they could be used to create
periodic arrays with attenuation bands due to resonance in the range
of frequencies below the BG of the array, i.e. in the range of low
frequencies. Thus, the resonance mechanism is used in our devices to
create attenuation peaks in the most difficult range of frequencies,
adding to both the scattering and the absorption effects.

In this Section, we analyze first the propagating properties of square periodic arrays ($a=0.33$ m)
of rigid SRR (RSRR) in order to analyze the behaviour of the resonances in a periodic system,
and afterwards we will analyze the combined effect scattering-resonances-absorption with absorbing SRR (ASRR). 
The size of the finite structures considered here is $4a\times 4a$
and the source is placed at the origin of coordinates. The SC has
been placed symmetrically to the source - receiver line at 1.5 m
away from the source. The IL has been calculated at a point located
3 m from the origin of coordinates in the direction of wave
propagation. The numerical predictions have been tested using
experimental measurements. The RSRR for the experimental set up have
been constructed from split ring tubes of PVC (rigid) cylinders.

\subsection{Rigid Split Ring Resonators (RSRR)}
\label{sec:Results1} The RSRR analyzed in this Section have been
designed with the following parameters: external radius $r=0.1$ m,
inner radius $r=0.095$ m and aperture width $L=0.02$ m. First, we
have analyzed the scattering problem of an isolated RSRR in order to
observe the behaviour of the resonant frequency.




A wave impinging the RSRR from the left is considered, presentting
the RSRR its aperture in this side, as one can see in the inset of
\ref{fig:Figure_4_def}.  The IL produced by an isolated RSRR has
been numerically obtained using FEM. \ref{fig:Figure_4_def}A shows a
clear resonance peak around 220 Hz. The open blue circles represent
the experimental measurements of the IL in good agreement with the
numerical predictions. Also, the localization of the pressure inside
the cavity for this resonant frequency can be observed in the inset
of \ref{fig:Figure_4_def}B. We note that, if the usual formula of
the 2D Helmholtz resonator
\footnote{$\nu_{Helmholtz}=\frac{c}{2\pi}\sqrt{\frac{A}{LS}}$, where
$A$ is the aperture, $L$ is the wall thickness and $S$ is the
surface of the cavity} is used for the RSRR presented here, the
first resonant mode should appear at a frequency of 610 Hz, which is
far away from the obtained in our analysis. This fact shows that the
considered RSRR does not behave as a 2D Helmholtz resonator.

\begin{figure}[hbt]
\begin{center}
\caption{ Acoustic properties of both a single and a periodic array
of  RSRR. (A) Numerical (blue continuous line) and experimental
(blue open circles) IL produced by an isolated RSRR. (C) Dispersion
relation of an square periodic array with $a=0.33$ m made of RSRR.
Strong grey area represents the attenuation band due to the
resonance and weak grey area represents the pseudogap at $\Gamma$X
direction. (B) and (D) are the numerical and experimental IL of a SC
made of RSRR respectively. Blue continuous and red dashed lines in
(B) (open blue circles and open red squares in (D)) are the IL in
the $\Gamma$X and $\Gamma$M directions respectively.}
\label{fig:Figure_4_def}
\end{center}
\end{figure}

The bands structure of the square array formed by the designed RSRR
have been also numerically calculated and represented in
\ref{fig:Figure_4_def}C. This diagram shows the existence of an
attenuation band in the range of low frequencies (around 220 Hz) due
to the resonance effect of the RSRR, independent of the BG of the
array (centered at 515 Hz). The numerical IL results are presented
in \ref{fig:Figure_4_def}B with the blue line (red dashed line)
representing the IL at $\Gamma$X ($\Gamma$M) direction. The
scattering problem reproduces both the first pseudogap at $\Gamma$X
direction and the resonance of the RSRR. On the other hand, note
that these results predict the non existence of a pseudogap at
$\Gamma$M direction. This fact shows the dependence of the array
effects on the incidence direction. However, one also can see that
the resonance effect is independent of the incidence direction: both
red (45$^\circ$) and blue (0$^\circ$) lines show the same peak at
low frequencies.

\ref{fig:Figure_4_def}D shows in blue open circles (red open
squares) the experimental measurements of the IL for the considered
device at $\Gamma$X ($\Gamma$M) direction. The good agreement with
the theoretical predictions allow us to validate the results and
demonstrates the appropriate choice of the theoretical tools to
analyze this problem.



\subsection{Absorbent Split Ring Resonators (ASRR)}
\label{sec:Results2} The last step of our study is the analysis of
the periodic distribution of absorbent split ring resonators (ASRR)
shown in \ref{fig:scatterer}. In this case the three attenuation
phenomena named at the beginning are considered. So, the scatterers
are RSRR covered with a layer of absorbent material with a thickness
$t=0.04$ m, and the designed device consist of a set of ASRR
arranged in the square array defined previously.

\begin{figure}[hbt]
\begin{center}
\caption{ Study of the acoustical properties of  a ASRR SC. (B)
Dispersion relation of an square periodic array with $a=0.33$ m made
of SRR with both $r=0.1$ m (black dashed lines) and $r=0.14$ m (red
continuous line). Strong grey area represents the attenuation band
due to the resonance, and weak grey area represents the pseudogap at
$\Gamma$X direction. (A) and (C) represent the numerical and
experimental IL of a SC made of ASRR respectively. Blue continuous
and red dashed lines in (A) (open blue circles and open red squares
in (C)) represent the IL at $\Gamma$X and $\Gamma$M directions
respectively. In all cases, the dot-dashed black line represents the
attenuation level predicted using Maekawa's model for a classical
barrier.} \label{fig:Figure_5_def}
\end{center}
\end{figure}

The study of the propagating properties of this SC is shown in
\ref{fig:Figure_5_def}. First of all, we have analyzed both the
array and the resonant effects using the PWE method.
\ref{fig:Figure_5_def}B shows the bands structure of the previous SC
made of RSRR with $r=0.1$ m (blue dotted lines). Although the bands
structure is only valid for the case of rigid cylinders (RSRR), we
have calculated the dispersion relation for the ASRR case (red
continuous line) supposing this device formed by rigid scatterers
($r=0.14$ m, being the inner radius $r=0.095$ m). This strategy
allow us to predict the variation of the attenuation produces in the
array due to an increasing of the radius of the scatterers. If we
compare both cases, the bands structure for the ASRR device predicts
a complete BG centered in 515 Hz, due to the increasing of the ff,
that not appears in the case of the RSRR SC. So, in the ASRR case
there is no propagation at any direction for this range of
frequencies.

On the other hand, as we have shown before, the resonant effect in
the RSRR SC introduces an attenuation band in the low frequency
range below the BG ($\nu=$220 Hz, see \ref{fig:Figure_4_def}C).
However, the corresponding attenuation band for the ASRR SC case
(red continuous line) showed in \ref{fig:Figure_5_def}B is shifted
in frequency to the lower frequencies. This interesting effect is
produced because the absorbent covering becomes part of the
resonator, increasing the wall thickness and producing a shifting of
the resonance frequency. This result can be used as a design tool
that could be exploited to attenuate other near ranges of
frequencies.

To determine the effect of the absorbent covering we have calculated
numerically the IL of the ASRR SC. Comparing the IL levels in
\ref{fig:Figure_5_def}A and \ref{fig:Figure_4_def}B, one can observe
that the average IL produced here is increased by the absorbing
covering practically in the whole range of frequencies and for both
$\Gamma$X and $\Gamma$M directions. Thus, the absorbing material
introduces a base line of attenuation independent of the angle of
incidence of the wave. Although the scattering problem has been
solved for the ASRR case and the bands structure is only valid for
the case of rigid scatterers, one can compare both results to
observe whether or not the absorbent covering destroys both the
scattering and the resonance effects. Thus, in figure
\ref{fig:Figure_5_def}A one can observe two attenuation peaks due to
both the multiple scattering at $\Gamma$X and $\Gamma$M directions
and the resonance effect over the base line of attenuation produced
by the absorbing covering. Both peaks have been predicted by the
bands structure. The experimental results are showed in
\ref{fig:Figure_5_def}C. One can observe again the good agreement
between the theoretical and experimental results.


Finally note that, as in the case of RSRR SC, the attenuation level
depends on the number of scatterers. We have observed that the
greater the number of rows, the higher IL. Obviously, this result is
in agreement with the mass law. However, it does not seem obvious
that both the multiple scattering and the resonance phenomena
continue to present the same properties as in the RSRR case when the
absorbent material is introduced. We have also observed that these
both attenuation effects are also increased with the number of ASRR.
Of course, the attenuation peak due to the resonant effect is
independent of the incident direction as in the RSRR case.

\section{Applications}
\label{sec:Applications} The IL of a SC made of ASRR is
characterized mainly by three acoustical properties: $i$) a high attenuation base line; 
$ii$) the structure preserve the properties of the periodicity,
meaning that, it preserves the BG although the absorbing covering is
surrounding the scatterers; $iii$) the resonances of each scatterer
are also preserved in the structure. Then, multiple scattering,
resonances and absorption co-exist in the same structure without
negative interference between them. Thus, the frequency response of
the system in the audible acoustic frequencies is similar to a
wideband bandstop filter that allows the transmission of some
frequency ranges but prevent others, attenuated to very low levels.

Sound manipulation is perhaps the most obvious application of the
structures presented in this work. In last years an increasing
interest in such a systems have motivated several works showing high
technological tools for such a SC as acoustic filters. One can built
natural SC made of periodic arrangement of trees \cite{Martinez06}
or even one can use the evanescent behaviour of waves inside the SC
in order to obtain an effective width to attenuate waves
\cite{Romero10a, Romero10b}. On the other hand, one can apply
optimization algorithms as for example evolutionary algorithms to
look for the best distribution of scatterers to attenuate predetermined frequencies
\cite{Romero09}. Taking into account that bandstop filters are
commonly used to eliminate particular frequencies of noise, we apply
this system to develop a device as an alternative to the
conventional acoustic barriers with several properties from the
acoustical point of view but also with additional aesthetic and
constructive characteristics.

\begin{figure}[hbt]
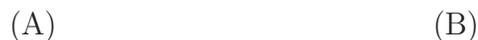

\begin{center}
(A)\hspace{5cm}(B) \caption{ (A) Picture of a
conventional acoustic barrier. (B) Simulated acoustic barrier based on SC}
\label{fig:Figure_6_def}
\end{center}
\end{figure}

The IL of the designed device is comparable with that predicted by
Maekawa for a classical barriers, as one can see in
\ref{fig:Figure_5_def}A and \ref{fig:Figure_5_def}C: there are
ranges of frequencies in which the SC produces better attenuation
and ranges in which the SC works worse than the classical barriers.
Thus, SC made of ASRR are suitable to be used as acoustic barriers
in certain ranges of frequencies. The high technology possibilities
because of the control of the multiple scattering, resonances and
absorption phenomena, as well as the non dependence of the IL with
the incident angle gives additional different properties than the
conventional acoustic barriers.

In \ref{fig:Figure_6_def} we show two pictures of both a
conventional acoustic barrier and a possible acoustic barrier based
on SC. Although the acoustic barrier based on SC could be wider and
more expensive (depending on the materials) than the conventional
barriers they can be attractive for some purposes because they are
transparent to water and wind, presenting tunable acoustical
properties and seeming aesthetically and constructively better than
the classical ones.

\section{Conclusions}
\label{sec:conclusions} Scatterers made of rigid walls presenting
resonant cavities and covered with absorbent materials are used in
this work to design a periodic distribution of ASRR. This SC shows a
constructive superposition of three physical phenomena being its attenuation properties tunable in
a wide band of frequencies by changing the parameters of the array,
the characteristics of the resonant cavity or the thickness and the
acoustical properties of the absorbent material. Multiple scattering
theory, plane wave expansion and finite element methods are used to
theoretically study the physical behaviour of the structure.
Experimental results obtained in an anechoic chamber are in good
agreement with both the analytical and numerical predictions. The
high possibilities to control the audible noise by means of
scattering, resonances and absorption give the possibility to design
a wideband bandstop acoustic filter specifically indicated to
attenuate audible noise. The designed structures produces
attenuation levels large enough to compete with the conventional
acoustic barriers introducing high technological procedures in their
design and opening the possibility to fabricate custom-tailored
bandstop acoustic filters based on ASRR SC.


\ack
This work was supported by MEC (Spanish Government) and FEDER
funds, under grands MAT2009-09438.


\section*{References}



\end{document}